\documentclass[onecolumn,showpacs,preprintnumbers,amsmath,amssymb,floatfix]{revtex4}

\usepackage{graphicx}
\usepackage{dcolumn}
\usepackage{bm}

\newcommand{\pom}{\tt I\! P}
\newcommand{\beq}{\begin{equation}}
\newcommand{\eeq}{\end{equation}}

\def \pom {{I\!\!P}}

\begin{document}

\title{Double $J/\psi$ production in central diffractive processes at the LHC}
\author{C. Brenner Mariotto$^{a}$ and V. P. Goncalves$^{b}$}

\affiliation{
$^a$ Instituto de Matem\'atica, Estat\'{\i}stica e F\'{\i}sica, Universidade Federal do Rio Grande\\
Av. It\'alia, km 8, Campus Carreiros, CEP 96203-900, Rio Grande, RS, Brazil\\
$^b$ High and Medium Energy Group, Instituto de F\'{\i}sica e Matem\'atica, Universidade Federal de Pelotas\\
Caixa Postal 354, CEP 96010-900, Pelotas, RS, Brazil
}

\begin{abstract}
In this paper we study  the double $J/\psi$ production in central diffractive processes considering the Resolved Pomeron model.  Based on the nonrelativistic QCD (NRQCD) factorization formalism  for the quarkonium production mechanism we estimate the rapidity and transverse momentum dependence of the cross section for the double $J/\psi$ production in  diffractive processes at LHC energies. The contributions of the color-singlet and color-octet channels are estimated and predictions for the total cross sections in the  kinematical regions of the LHC experiments are also presented. Our results demonstrate  that the   contribution of  central diffractive processes is not negligible and that its study can be useful to test the Resolved Pomeron model.
\end{abstract}


\maketitle


The study of the production of heavy quarkonium states provides a unique laboratory in which one can explore the interplay between perturbative and nonperturbative effects in QCD (For a  review see, e.g., Ref. \cite{review_nrqcd}).
In the last years, several theoretical approaches have been proposed for the calculation of these states, as for instance, the Color Singlet model (CSM), the color evaporation model, the Non Relativistic QCD (NRQCD)  approach, the fragmentation approach,  and the $k_T$-factorization approach. Albeit  considerable efforts both in theory and experiments, the quarkonium production mechanism is still not fully understood. A particular example is the  double quarkonium production in inclusive processes, i. e. processes where the two incident hadrons dissociate in the interaction. In the last years, the measurements reported by the LHCb \cite{lhcb_inc}, CMS \cite{cms} and D0 \cite{d0} collaborations  at the LHC and the Tevatron have posed significant challenges to our understanding of the quarkonium production.  Currently, there are in the literature several predictions for the total cross section and differential distributions \cite{duplo1,duplo2,qiao,duplo3,duplo4,ko,duplo5,duplo6,duplo7,duplo8,lansberg_nlo1,nrqcd_nlo,lansberg_nlo2}, at leading-order and next-to-leading order of the perturbative expansion, some of them  using the Color Singlet model and others using the NRQCD formalism.  
The results of these studies indicate that the discrimination between the different approaches in inclusive processes will be a hard task, since double parton scattering (DPS) processes are expected to contribute at high energies and the contribution of this new mechanism is still an open question \cite{dps1,dps2,dps3}. In contrast, the DPS contribution for diffractive processes, where the incident hadrons remain intact, is expected to be negligible \cite{kmr_duplo}. This motivates  the study of the double quarkonium production in  diffractive interactions in order to test the production mechanism.

In recent years the diffractive processes have attracted much attention as a way of amplifying the physics programme at hadronic colliders, including searching for New Physics (For a  review see, e.g. Ref. \cite{forshaw}). The investigation of these reactions at high energies gives important information about the structure of hadrons and their interaction mechanisms. 
The diffractive physics has been tested in hadron-hadron collisions considering different theoretical approaches and distinct final states like dijets, electroweak vector bosons, dileptons, heavy quarks, quarkonium + photon and jet + photon (See, e.g., Refs. \cite{Covolan:2002kh,roman,golec2,MMM1,ingelman,golec,quark_photon,schurek,Brodsky:2006wb,Kopeliovich:2006tk,Kopeliovich:2007vs,Pasechnik:2011nw, Kepka:2010hu,Marquet:2012ra,vic_cris,royon}). One of these approaches is the Resolved Pomeron Model,  proposed by Ingelman and Schlein in Ref. \cite{IS}, which assumes the validity of the diffractive factorization formalism and that the Pomeron has a partonic structure. The basic idea is that the hard scattering resolves the quark and gluon content in the Pomeron \cite{IS}, which can be obtained by analysing the experimental data from diffractive deep inelastic scattering (DDIS) at HERA, providing us with the diffractive distributions of singlet quarks and gluons in the Pomeron \cite{H1diff}. However, other approaches based on very distinct assumptions are also able to describe the current scarce experimental data. Consequently, the present scenario for diffractive processes is unclear, motivating the study of alternative processes which could allow to constrain the correct description of the Pomeron.

In this paper we propose the study of the double $J/\Psi$ production  as a complementary test of diffractive processes and the pomeron structure. In particular, we present a detailed analysis of the rapidity and transverse momentum dependence of the cross section for the double $J/\Psi$ production in diffractive  processes considering the nonrelativistic QCD (NRQCD) factorization formalism \cite{nrqcd}  for the quarkonium production mechanism. Moreover, a comparison with the inclusive production is presented. Our analysis is strongly motivated  by the recent  LHCb data \cite{lhcb_dif} for the diffractive production of this final state and by the recent theoretical study performed in Ref. \cite{kmr_duplo}, which have studied the double $J/\Psi$ production in exclusive processes. Our goal is to present a complementary analysis using a distinct model for the treatment of the diffractive interactions.

\begin{figure}[t]
\begin{center}
\scalebox{0.35}{\includegraphics{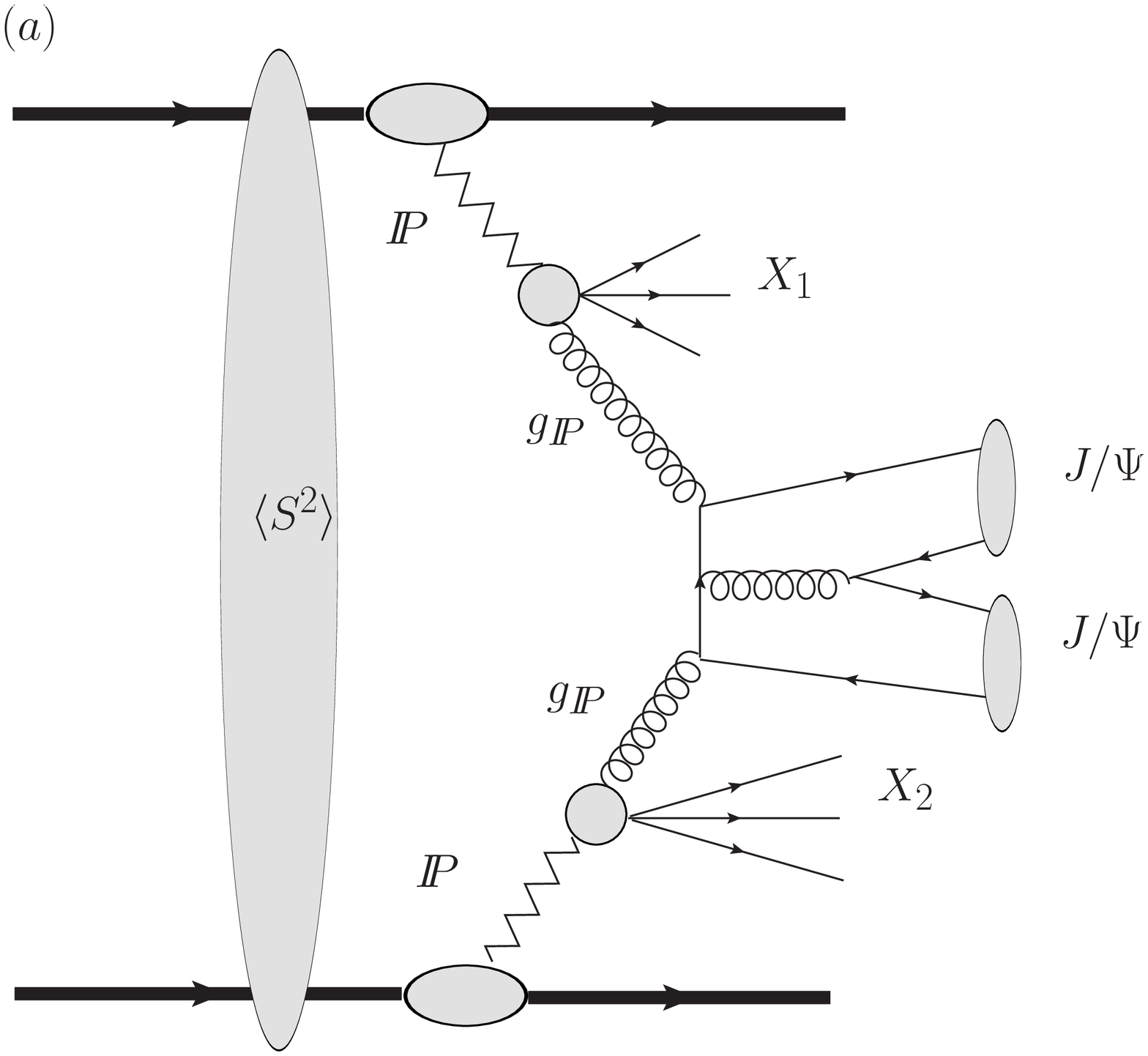}}
\scalebox{0.35}{\includegraphics{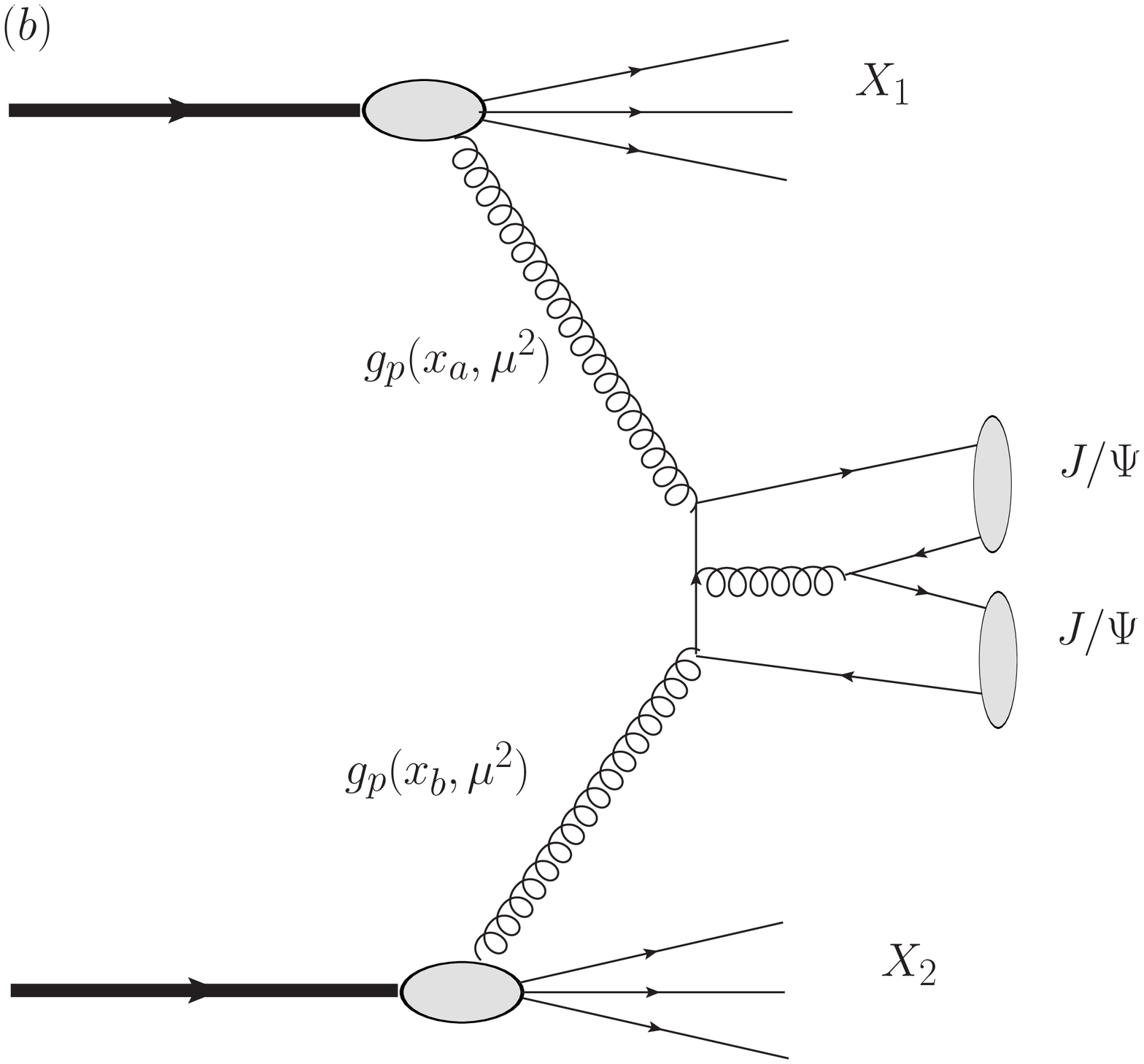}}
\caption{Typical diagrams for the double $J/\psi$ production in (a)  central diffractive and (b) inclusive processes.}
\label{centraldif}
\end{center}
\end{figure}

In the following we apply the Resolved Pomeron Model \cite{IS} for the central diffractive double $J/\psi$ production. This model assumes that the Pomeron has a well defined  partonic structure and that the hard process takes place in Pomeron - Pomeron processes. The contributing diagrams are the same as in the inclusive case, and the $J/\psi$ production in the central diffractive processes is described by diagrams like those ilustrated in Fig. \ref{centraldif} (a), with ${I\!\!P} {I\!\!P} $ interactions.
Differently from the inclusive processes illustrated in Fig. \ref{centraldif} (b), the diffractive processes are characterized 
by the presence of one or two intact very forward hadrons and empty regions in pseudo-rapidity, called rapidity gaps, in the final state. In this paper we restrict our study to the central diffractive processes, which are characterized by two intact hadrons, two rapidity gaps  and 
soft particles accompanying the double $J/\Psi$ production [See Fig. \ref{centraldif} (a)]. The presence of these soft particles associated to the Pomeron remnants is a characteristic of the Resolved Pomeron model. In contrast, in the case of the  central exclusive production of two $J/\Psi$ discussed in Ref. \cite{kmr_duplo},  nothing else is produced except the leading hadrons and the central object.

In the NRQCD formalism \cite{nrqcd} the cross section for the diffractive production of a heavy quarkonium pair can be factorized as follows
\begin{eqnarray}
d\sigma (pp\rightarrow p \otimes X_1 + H_1 + H_2 + X_2 \otimes p) = \langle S^2 \rangle \sum_{a,b,n_1,n_2}f^D_{a/p} \otimes f^D_{b/p} \otimes d\hat{\sigma}[ab\to Q\bar{Q}[n_1]+Q\bar{Q}[n_2]+X] \cdot \langle {\cal{O}}^{H_1}_{n_1} \rangle \langle {\cal{O}}^{H_2}_{n_2} \rangle \,\,,
\end{eqnarray}
 where $\otimes$ represents the presence of a rapidity gap in the final state, $X_i$ the remnants of the Pomeron, $\langle {\mathcal S}^2 \rangle$ is the gap survival probability (see below), $f^D$ are the diffractive parton distributions and the coefficients $d\hat{\sigma}[ab\to Q\bar{Q}[n_1]+Q\bar{Q}[n_2]+X]$
are perturbatively calculable short distance cross sections for the production of the two heavy quark pairs in an intermediate Fock state $n_i={}^{2S+1}L_J^{[i]}$ ($i=1,8$), which does not have to be color neutral. The $\langle {\cal{O}}^{H_i}_{n_i}\rangle$
are nonperturbative long distance matrix elements (LDME), which describe the transition of the intermediate $Q\bar{Q}$ in the physical state $H$ via soft gluon radiation. Currently, these elements have to be extracted from global fits to quarkonium data as performed, for instance, in Ref. \cite{bute}. 
In the Color Singlet Model \cite{csm}, only the states with the same quantum numbers as the resulting quarkonium contribute to the formation of a bound $Q\bar{Q}$ state.  In contrast, in NRQCD, also color octet $Q\bar{Q}$ states contribute to the quarkonium production cross section via soft gluon radiation. The Color Singlet Model can be obtained from NRQCD factorization by retaining, for a given process, only the contribution that is associated with the color-singlet LDME of the lowest non-trivial order in $v$, which is the typical velocity of the heavy quark or antiquark in the quarkonium rest frame. 
At high energies the double quarkonium production is dominated by gluon - gluon interactions. Consequently, 
the differential cross section for the double $J/\psi$ production in central diffractive (CD) processes can be written as
 \begin{eqnarray}
\frac{d\sigma^{CD} }{dydp_T^2}=  \langle S^2 \rangle \cdot \int_{x_{a\, min}} dx_a
{g^D(x_a,\mu^2)g^D(x_b,\mu^2)}\frac{x_ax_b}{2x_a-\bar{x}_Te^y}
\sum_{i=1,8} \frac{d\hat{\sigma}}{d\hat{t}}[gg\rightarrow
2 c\bar{c}_i(^3S_1)]
\cdot \langle {\cal{O}}_i^{J/\psi}(^3S_1) \rangle^2  \,\,,
\label{csdif}
\end{eqnarray}
where 
$x_{a\, min}=\frac{\bar{x}_Te^{y}}{2-\bar{x}_Te^{-y} }$, 
$x_b=\frac{x_a\bar{x}_Te^{-y}}{2x_a-\bar{x}_Te^y}$, 
$\bar{x}_T=\frac{2m_T}{\sqrt{s}}$ and $m_T=\sqrt{M^2+p_T^2}$. Here  $M$ is the $J/\psi$ mass, $p_T$ its transverse momentum and $y$ its rapidity. The $J/\psi$ transverse mass is taken as the hard scale of the problem, with $\mu_R=\mu_F=m_T$.
Here, ${g^D(x_i,\mu^2)}$  are the diffractive gluon distribution functions from the two colliding protons.  
In the present work, the diffractive gluon distributions in the proton are taken from the Resolved Pomeron Model \cite{IS}, where they are defined as a convolution of the Pomeron flux emitted by the proton, $f_{I\!\!P}(x_{I\!\!P})$, and the gluon distribution in the Pomeron, $g_{I\!\!P}(\beta, \mu^2)$, where $\beta$ is the momentum fraction carried by the partons inside the Pomeron. The Pomeron flux is given by $f_{I\!\!P}(x_{I\!\!P})= \int_{t_{min}}^{t_{max}} dt f_{\pom/p}(x_{{I\!\!P}}, t)$, where $f_{\pom/p}(x_{\pom}, t) = A_{\pom} \cdot \frac{e^{B_{\pom} t}}{x_{\pom}^{2\alpha_{\pom} (t)-1}}$ and $t_{min}$, $t_{max}$ are kinematic boundaries. The Pomeron flux factor is motivated by Regge theory, where the Pomeron trajectory assumed to be linear, $\alpha_{\pom} (t)= \alpha_{\pom} (0) + \alpha_{\pom}^\prime t$, and the parameters $B_{\pom}$, $\alpha_{\pom}^\prime$ and their uncertainties are obtained from fits to H1 data  \cite{H1diff}. The diffractive  gluon distribution is then given by
\begin{eqnarray}
{ g^D(x,\mu^2)}=\int dx_{I\!\!P}d\beta \delta (x-x_{I\!\!P}\beta)f_{I\!\!P}(x_{I\!\!P})g_{I\!\!P}(\beta, \mu^2)={ \int_x^1 \frac{dx_{I\!\!P}}{x_{I\!\!P}} f_{I\!\!P}(x_{I\!\!P}) g_{I\!\!P}\left(\frac{x}{x_{I\!\!P}}, \mu^2\right)}  \,\,.
\end{eqnarray}
In our analysis we use the diffractive gluon distribution obtained by the H1 Collaboration at DESY-HERA \cite{H1diff}. It is important to emphasize that the cross section for the diffractive production of two $J/\Psi$  is strongly sensitive to the Pomeron structure due to the quadratic dependence on the diffractive gluon distribution. 
Finally, $\frac{d\hat{\sigma}}{d\hat{t}}$ in Eq. (\ref{csdif}) are the hard scattering differential cross sections, which we assume to be given by leading order (LO) $\alpha_s^4$ expressions. As demonstrated in Ref. \cite{ko} the corresponding Feynman diagrams can be classified into two groups: (a) diagrams associated to the nonfragmentation contribution, composed by 31 Feynman diagrams, with the leading contribution being the color singlet $(c\bar{c})_1({}^3S_1) + (c\bar{c})_1({}^3S_1)$ channel, and (b) diagrams associated to the gluon fragmentation contribution, composed by 72 Feynman diagrams, with the main contribution associated to the color octet 
$(c\bar{c})_8({}^3S_1) + (c\bar{c})_8({}^3S_1)$ channel. 
For the gluon-initiated color singlet contributions, one has \cite{qiao}
\begin{eqnarray}
\frac{d\hat{\sigma}}{d\hat{t}}[gg\rightarrow 2 c\bar{c}_1({}^3S_1)]
\cdot \langle O_1^{J/\psi}({}^3S_1) \rangle^2 =
\frac{16\pi \alpha_s^4  |R(0)|^4}{81 M^2s^8(M^2-t)^4(M^2-u)^4}\sum_{jkl} a_{jkl}M^jt^ku^l 
\label{csmdsdt}
\end{eqnarray}
where, as in the Ref. \cite{qiao},  $|R(0)|^2=0.8\,$ GeV$^3$ is the squared radial function at the origin, $s$, $t$ and $u$ are the usual Mandelstam variables, $M=2m_c$, and $m_c=1.5\,$ GeV. The detailed expressions for the $a_{jkl}$ coefficients in Eq. (\ref{csmdsdt}) are given in Ref. \cite{qiao}.
On the other hand, for the color octet contributions, the differential cross section for the gluon initiated partonic subprocesses was calculated in Ref. \cite{ko} and can be written as  
\begin{eqnarray}
\frac{d\hat{\sigma}}{d\hat{t}}[gg\rightarrow 2 c\bar{c}_8({}^3S_1)]
 =
\frac{\pi \alpha_s^4 }{972 M^6 s^8(M^2-t)^4(M^2-u)^4}\sum_{j=0}^{14} a_j M^{2j} \,, 
\label{comdsdt}
\end{eqnarray}
where the $a_j$ coefficients can be found in Ref. \cite{ko}. In the considered case, the only relevant NRQCD matrix element is $\langle O_8^{J/\psi}({}^3S_1) \rangle=3.9\times 10^{-3} GeV^3$, taken from \cite{Braaten2000}. This value has been updated in a recent fit to world data \cite{bute} which gives $\langle O_8^{J/\psi}({}^3S_1) \rangle=1.68\times 10^{-3} GeV^3$. Using this new value, our results for the COM contributions would decrease by a factor $2.3$, which is inside the uncertainties of our results. { Moreover, it is important to emphasize that, as in Ref.  \cite{ko}, we have disregarded the contributions from the $\langle O_8^{J/\psi}({}^1S_0)\rangle$ transition. This approximation is reasonable at high transverse momentum $p_T$. In contrast, for small values of $p_T$, these contributions are not negligible, dominating the COM contributions, as demonstrated in Refs.  \cite{Braaten2000,bute}. Consequently, our COM predictions, denoted COM$^*$ in what follows, should be considered an incomplete evaluation of the COM contributions at low-$p_T$. However, it is important to emphasize that the inclusion of the $\langle O_8^{J/\psi}({}^1S_0)\rangle$ transition in our calculations would not modify our main conclusions, since the CSM contribution largely dominates the cross section at small values of the transverse momentum (see below).}

\begin{figure}[t]
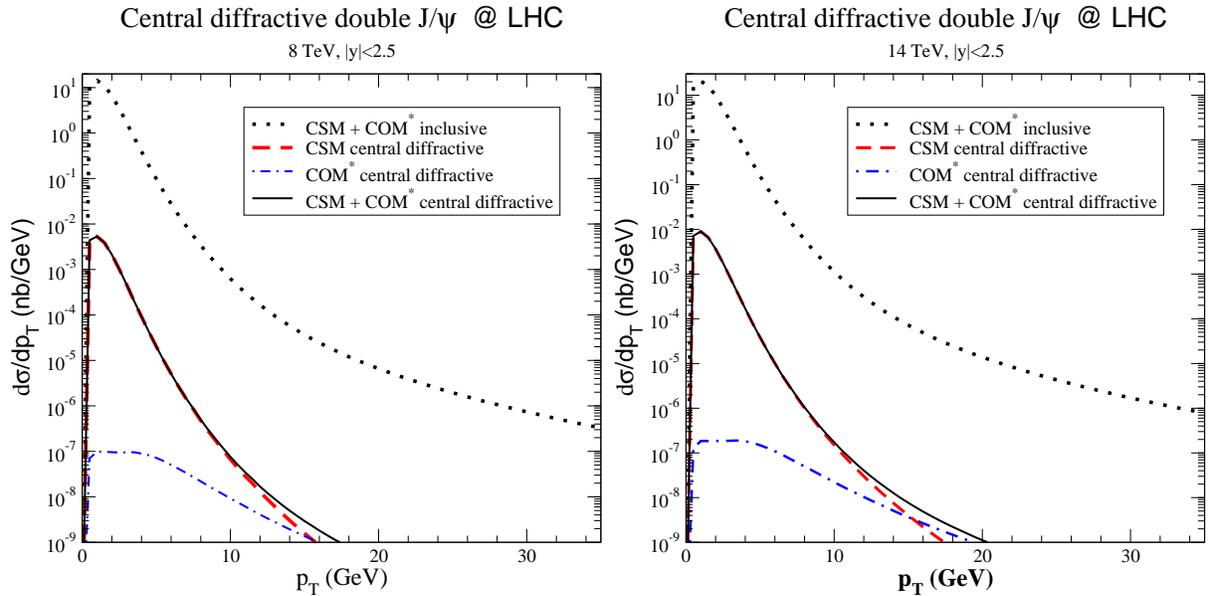

\begin{center}
\scalebox{0.55}{\includegraphics{doublejpsilhc8tev2.eps}}
\scalebox{0.55}{\includegraphics{doublejpsilhc14tev2.eps}}
\caption{(Color online) Transverse momentum distributions for double $J/\psi$ production in central diffractive processes at $\sqrt{s} = 8$ TeV (left panel) and 14 TeV (right panel). The prediction for the inclusive production is presented for comparison.}
\label{centraldifdoublept}
\end{center}
\end{figure}

In order to obtain realistic predictions for the double $J/\Psi$ production in central diffractive processes, it is crucial to use an adequate value for the gap survival probability, $\langle {\mathcal S}^2 \rangle$. This factor is the probability that secondaries, which are produced by soft rescatterings do not populate the rapidity gaps, and depends on the particles involved in the process and  in the center-of-mass energy.  In what follows, following Ref. \cite{kmr}, we will assume  that $\langle {\mathcal S}^2 \rangle = 2 \, \%$ for proton - proton collisions at LHC  energies. However, this subject deserves a more detailed analysis, since the magnitude of $\langle {\mathcal S}^2 \rangle$ is still a theme of intense debate in the literature (See, e.g., Ref. \cite{maor}).

In Fig. \ref{centraldifdoublept} we present our results for the transverse momentum distribution for the double $J/\Psi$ production at midrapidities ($|y|\le 2.5$) in central diffractive processes for LHC energies. For the sake of comparison, we also show the results for the inclusive case, obtained using the CTEQ6L parametrization \cite{cteq} for the gluon distribution in the proton, which agree with the predictions presented in Ref. \cite{ko}.
The central diffractive predictions at small $p_T$ are a factor $\approx 10^3$ smaller than in the inclusive one. We obtain that the $p_T$ distributions in the low $p_T$ region are dominated by the color singlet contributions. Moreover, as in the inclusive case \cite{ko}, the distribution vanishes at $p_T = 0$ and increases rapidly until it reaches a maximum at $p_T \approx 1.5$ GeV. Then it decreases monotonically as $p_T$ increases. We obtain that {\it at leading order}  the color singlet  contributions are dominant except at large $p_T$. The cross-over, beyond that the color octet contributions start to be dominant, occurs at  $p_T \approx 15$ GeV. Furthermore, in the dominant low-$p_T$ peak, the color octet contribution is four orders of magnitude less important than the color singlet one. It is important to emphasize that the recent studies of the next-to-leading order corrections for the inclusive double $J/\Psi$ production performed in Ref. \cite{lansberg_nlo1} indicate the the color singlet contributions also are dominant at large-$p_T$. A similar behaviour also is expected in the case of the double $J/\Psi$ production in central diffractive  processes if the NLO corrections are taken into account.

\begin{figure}[t]
\begin{center}
\scalebox{0.55}{\includegraphics{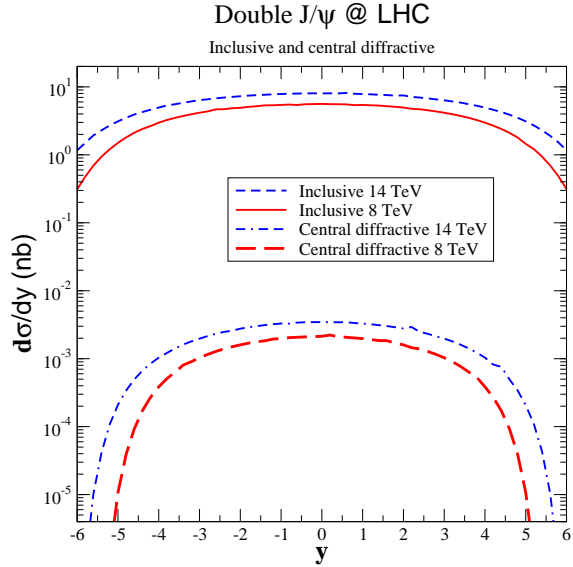}}
\caption{(Color online) Rapidity distributions for double $J/\psi$ production in inclusive and central diffractive processes in  $pp$ collisions at $\sqrt{s} = 8$  and 14 TeV.}
\label{centraldifdoubley}
\end{center}
\end{figure}

In Fig. \ref{centraldifdoubley} we present our results for the rapidity distribution for the double $J/\psi$ production in central diffractive processes at $\sqrt{s} = 8$ and 14 TeV. The inclusive predictions are also presented for comparison. As  expected our predictions increase with the energy.  We have verified that the color singlet contributions dominate all regions for the $p_T$-integrated spectra, with the  color octet one being  negligible in all rapidity regions. Moreover, we have a reduction of four orders of magnitude when going from the inclusive to the central diffractive case.  Our results for the rapidity distribution allows us to obtain the predictions for the  cross section in distinct rapidity ranges covered by the different LHC experiments.
In Table \ref{tab1} we present our predictions for the total cross section considering the production at midrapidities ($|y|\le 2.5$) , which can be analyzed by CMS, ATLAS and ALICE Collaborations, and for the kinematical range probed by the LHCb Collaboration ($2  \le y \le 4.5$). For comparison we present the predictions for the inclusive production and  for the double $J/\Psi$ production  in central exclusive processes (CEP) obtained in Ref. \cite{kmr_duplo} using the   CTEQ6L parton distribution functions. { In the case of midrapidities, we also present our predictions considering a cutoff in the  transverse momentum ($p_T \ge 6$ GeV), since the CMS and ATLAS detectors have a very low acceptance at low $p_T$. Our predictions for the inclusive production agree with those presented in Ref. \cite{ko} for the integrated cross sections obtained without a cutoff in the transverse momentum. If the cutoff is taken into account, the predictions for the double $J/\Psi$ production at midrapidities are reduced by three orders of magnitude. }
 We obtain that our predictions for the central diffractive production are similar to those for the central exclusive production obtained in Ref. \cite{kmr_duplo}. As already emphasized above, 
the topology of the final state of these two processes is different. While in central exclusive processes one has only the leading hadrons, two $J/\psi$'s and nothing else, in the central diffractive case one expect to have some extra particles coming from the Pomeron remnants. Consequently, in principle, the experimental separation between these two processes can be performed at { smaller} luminosities, as those presented in the LHCb analysis \cite{lhcb_dif}. However,  
it is not obvious if the double diffractive and the central exclusive mechanisms could be differentiated experimentally at the LHC in the next run. Unfortunately, due to the high luminosity and large pile-up environment the separation of the diffractive processes considering the rapidity gaps and the detection or not of the remnants of the Pomeron will be a hard task. The identification of diffractive processes should occur by tagging the intact protons in the final state using forward detectors to be installed at LHC. In this case, both contributions of the central diffractive and central exclusive processes for the double $J/\Psi$ should be taken into account.

\begin{table}[t]
\begin{center}
\scriptsize
\begin{tabular}{|c|c|c|c|c|}
\hline
\hline
CM energy & Kinematical ranges &  Inclusive &  Central Diffractive  & {CEP } \\
\hline
\hline
 8 TeV & $|y|<2.5$ &  27203 pb  &  9.51 pb  & {10 pb}\\
 \hline
 8 TeV & $|y|<2.5$ and $p_T \ge 6.0$ GeV &  29 pb  &  $4.5 \times 10^{-3}$ pb  & --\\
 \hline
8 TeV & $2<y<4.5$ &  9709 pb  &  2.16 pb  & {2.5 pb}\\
\hline
\hline
 14 TeV& $|y|<2.5$ &  39690 pb & 16.02  pb & {17 pb} \\
\hline
 14 TeV & $|y|<2.5$ and $p_T \ge 6.0$ GeV &  46 pb  &  $9.2 \times 10^{-3}$ pb  & --\\
\hline
14 TeV & $2<y<4.5$ &  15220 pb & 4.43  pb  & {4.7 pb} \\
  \hline
\hline
\end{tabular}
\caption{Total cross sections  for double $J/\psi$ production for different energies and distinct rapidity and transverse momentum cuts. Also shown the predictions  from Ref. \cite{kmr_duplo} for the central exclusive production (CEP).}
\label{tab1}
\end{center}
\end{table}

Some comments about our predictions are in order. Firstly, in our estimates we are not including the feed-down from excited states, like $J/\Psi + \Psi^{\prime}$, and relativistic corrections. Previous studies indicate that feed-down can significantly contribute for the double $J/\Psi$ cross section \cite{lansberg_nlo2}, while relativistic corrections decrease the magnitude of the cross section \cite{duplo6,duplo7}. Secondly, as the cross section is proportional to $\alpha_s^4$ and our calculations have been performed at leading order, our final results strongly depend on the choice of the hard scale. Clearly, the treatment of the central diffractive process considering the partonic cross sections at next-to-leading order is an important task for the future. However, the calculation of these corrections is still in progress \cite{lansberg_nlo1,nrqcd_nlo,lansberg_nlo2}.

Finally, lets summarize our main results and conclusions. Studies of the double $J/\Psi$ production are expected to provide important insights for improving the theoretical description of the quarkonium production mechanism. In the last years, several groups have discussed in detail the production of this final state in inclusive processes. However, as recently demonstrated by the LHCb Collaboration, the study of the double $J/\Psi$ production in diffractive processes is feasible. 
It has motivated the analysis of this process in the framework of the Resolved Pomeron model, which is one the the current models for the Pomeron. In this paper we have estimated the momentum and rapidity distributions for the double $J/\Psi$ production in central diffractive processes at LHC energies. Our results indicate that the contribution of the central diffractive processes is not negligible and that its study can be useful to test the Resolved Pomeron model.
In the future, we plan to extend our analysis for other heavy vector mesons in the final state as well as for single diffractive processes.

\begin{acknowledgments}
 This research was supported by CNPq, CAPES and FAPERGS, Brazil. 
\end{acknowledgments}

\end{document}